# Effects of anisotropic in-plane strains on the phase diagram of $Ba_xSr_{1-x}TiO_3$ thin film

Y. M. Tao*, Y. Z. Wu

Jiangsu Laboratory of Advanced Functional Materials, Department of Physics, Changshu Institute of Technology, Changshu 215500, P. R. China

**Abstract**

Based on Landau-Devonshire (LD) phenomenological theory, phase diagram of epitaxial BST50/50 thin films on anisotropic in-plane strains is investigated. Different from $BaTiO_3$ thin films, the paraelectric phase appears under the anisotropic misfit strains on BST50/50 thin films at the room temperature. The pyroelectric property of the BST films is also calculated, we find that the position of pyroelectric peak greatly depends on anisotropic misfit strains.

---

* Corresponding author.
   E-mail address: ymt@cslg.edu.cn (Y. M. Tao).

# I. Introduction

Perovskite barium strontium titanate ($Ba_xSr_{1-x}TiO_3$, or BST) that replaces lead titanate is a kind of "cleaner" material. Due to its high dielectric constant, low dielectric loss, low leakage current, large dielectric breakdown strength, low temperature coefficient for dielectric constant and composition-dependent Curie temperature, $Ba_xSr_{1-x}TiO_3$ thin films [1] have been investigated intensively for applications in dynamical random access memory (DRAM), tunable microwave devices, by-pass capacitors and infrared detectors [2, 3]. $BaTiO_3$ is a prototype ferroelectric for which the origin of ferroelectricity derives from the displacement of ions relative to each other, while $SrTiO_3$ is quantum paraelectric or incipient ferroelectric [4]. The ferroelectric and dielectric properties of BST thin films vary with Ba and Sr compositions. Especially, Curie temperature of $Ba_{0.5}Sr_{0.5}TiO_3$ below operating temperature is used for practical device applications. Since paraelectric state of ferroelectric materials has lower dielectric loss due to the disappearance of hysteresis, compositional barium strontium titanate has been widely used for room temperature applications [5-8].

Internal stress that originates from mismatch of the lattice parameters between film and substrate has a crucial impact on the quality of epitaxial ferroelectric films materials. With different substrate materials, there will be different misfit strains. Previous theories and experiments have investigated the strain effects of ferroelectric thin films grown on (001)-oriented cubic substrates. In those studies, the in-plane strain was supposed to be isotropic in the film. N. A. Pertsev et al. studied such a case and gave the misfit-temperature phase diagrams of single domain $BaTiO_3$ and $PbTiO_3$ films with a phenomenological thermodynamic theory [9]. Z. G. Ban et al. used phenomenological model to study dielectric response, pyroelectric response and tunability of epitaxial BST films on "compressive" $LaAlO_3$ substrate and "tensile" $MgO$ substrate [10-12].

Recently, A. G. Zembilgotov et al. investigate $PbTiO_3$ and PST films on orthorhombic substrates that induce the in-plane strain anisotropy [13,14]. The misfit strain along a crystalline axis differs from the misfit strain along the other crystalline axis. In a series of experiments, Simon et al. have analyzed the structure, anisotropic strain, dielectric response and tunability of (110)-oriented BST60/40 thin films grown on orthorhombic (100)-oriented $NdGaO_3$ [15-18]. Theoretically, (110) BST60/40 films on (001) orthorhombic NGO substrates have been analyzed [19], the results show significant improvement in the tunability along both in-plane and out-of-plane directions of BST60/40 films with increasing film thickness compared to the similar films on isotropic cubic substrates. Therefore, the study of anisotropic misfit strains on BST thin films has great significance.

The (110) film grown on (001) substrate or (100) film grown on (110) substrate both lead to the difference of stress on x-axis and y-axis. They have the in-plane anisotropic strains which are caused by the film and the substrate respectively. In this paper, we use LD phenomenological model to investigate the effects of anisotropic misfit strains on the physical properties of $Ba_xSr_{1-x}TiO_3$ films. We define anisotropic misfit strains as $e_1 = (b_1 - a_1)/b_1$ in x-axis and $e_2 = (b_2 - a_2)/b_2$ in y-axis, where $b_i$ ($i = 1, 2$) is the substrate lattice parameter and $a_i$ ($i = 1, 2$) is the lattice constant of the film at the stress-free state. A quantitative estimation on the dielectric response and pyroelectric response as a function of the anisotropic misfit strains for BST compositional film is presented.

**II. Theory**

The thermodynamic potential $F$ of BST film can be expressed in terms of the polarization $P_i$, applied field $E_i$ and anisotropic misfit strains, it is given by [9-20]

$$F = \alpha_1^* P_1^2 + \alpha_2^* P_2^2 + \alpha_3^* P_3^2 + \alpha_{11}^* (P_1^4 + P_2^4) + \alpha_{33}^* P_3^4 + \alpha_{12}^* P_1^2 P_2^2$$
$$+ \alpha_{13}^* (P_2^2 P_3^2 + P_1^2 P_3^2) + \alpha_{111} (P_1^6 + P_2^6 + P_3^6)$$
$$+ \alpha_{112} [(P_1^4 (P_2^2 + P_3^2) + P_2^4 (P_1^2 + P_3^2) + P_3^4 (P_1^2 + P_2^2))]$$
$$+ \alpha_{123} P_1^2 P_2^2 P_3^2 + \frac{S_{11}(e_1^2 + e_2^2) - 2 S_{12} e_1 e_2}{2(s_{11}^2 - s_{12}^2)} - E_1 P_1 - E_2 P_2 - E_3 P_3. \quad (1)$$

The vector and tensor quantities are defined in Cartesian coordinate system with $P_1 \| [100]$, $P_2 \| [010]$, and $P_3 \| [001]$. The renormalized coefficients in the free energy expression Eq. (1) are

$$\alpha_1^* = \alpha_1 - \frac{Q_{12}(s_{11}e_2 - s_{12}e_1) + Q_{11}(s_{11}e_1 - s_{12}e_2)}{s_{11}^2 - s_{12}^2}, \quad (2)$$

$$\alpha_2^* = \alpha_1 - \frac{Q_{12}(s_{11}e_1 - s_{12}e_2) + Q_{11}(s_{11}e_2 - s_{12}e_1)}{s_{11}^2 - s_{12}^2}, \quad (3)$$

$$\alpha_3^* = \alpha_1 - \frac{Q_{12}(e_1 + e_2)}{s_{11} + s_{12}}, \quad (4)$$

$$a_{11}^* = a_{11} + \frac{1}{2} \frac{1}{s_{11}^2 - s_{12}^2} [(Q_{11}^2 + Q_{12}^2)s_{11} - 2Q_{11}Q_{12}s_{12}], \quad (5)$$

$$a_{33}^* = a_{11} + \frac{Q_{12}^2}{s_{11} + s_{12}}, \quad (6)$$

$$a_{12}^* = a_{12} - \frac{1}{s_{11}^2 - s_{12}^2} [(Q_{11}^2 + Q_{12}^2)s_{12} - 2Q_{11}Q_{12}s_{11}] + \frac{Q_{44}^2}{2s_{44}}, \quad (7)$$

$$a_{13}^* = a_{12} + \frac{Q_{12}(Q_{11} + Q_{12})}{s_{11} + s_{12}}, \quad (8)$$

where $a_1$ is the dielectric stiffness, $a_{ij}$ and $a_{ijk}$ are higher order stiffness coefficients at constant stress, $Q_{ij}$ are the electrostrictive coefficients, and $s_{ij}$ are the elastic compliances of the film. The temperature dependence of the dielectric stiffness $a_1$ is given by the Curie-Weiss law, $a_1 = (T - T_0)/2\varepsilon_0 C$, where $T_0$ and $C$ are the Curie-Weiss temperature and constant of bulk material, respectively, and $\varepsilon_0$ is the permittivity of free space. The parameters used for the calculation of the renormalized coefficients for BST50/50 films are given in Table I [12], where the contribution of sixth-order polarization terms to the free energy is neglected.

Based on Eqs. from (1) to (8), we take the minima of $F$ with respect to the components of the polarization for all possible phases in epitaxial films as identified by Wang et al. [13]: (ⅰ) the $a_1$ phase, where $P_1 \neq 0$ and $P_2 = P_3 = 0$; (ⅱ) the $a_2$ phase, where $P_2 \neq 0$ and $P_1 = P_3 = 0$; (ⅲ) the $c$ phase, where $P_3 \neq 0$ and $P_1 = P_2 = 0$; (ⅳ) the $a_1c$ phase, where $P_1 \neq 0$, $P_3 \neq 0$ and $P_2 = 0$; (ⅴ) the $a_2c$ phase, where $P_2 \neq 0$, $P_3 \neq 0$ and $P_1 = 0$; (ⅵ) the $a_1a_2$ phase, where $P_1 \neq 0$, $P_2 \neq 0$ and $P_3 = 0$; (ⅶ) the $\gamma$ phase, where $P_1 \neq 0$, $P_2 \neq 0$ and $P_3 \neq 0$ and (ⅷ) the paraelectric phase, where $P_1 = 0$, $P_2 = 0$ and $P_3 = 0$.

The components of the spontaneous polarizations are given by the equations $\dfrac{\partial F}{\partial P_i} = 0$, i.e.

$$\frac{\partial F}{\partial P_1} = 2(a_1^* + a_{13}^* P_3^2 + a_{12}^* P_2^2)P_1 + 4a_{11}^* P_1^3 - E_1 = 0, \tag{9}$$

$$\frac{\partial F}{\partial P_2} = 2(a_2^* + a_{13}^* P_3^2 + a_{12}^* P_1^2)P_2 + 4a_{11}^* P_2^3 - E_2 = 0, \tag{10}$$

$$\frac{\partial F}{\partial P_3} = 2(a_3^* + a_{13}^* (P_1^2 + P_2^2))P_3 + 4a_{33}^* P_3^3 - E_3 = 0. \tag{11}$$

The electric field dependent dielectric constants along $[100]$, $[010]$, and $[001]$ directions are written as following,

$$\varepsilon_{11} = (\frac{\partial^2 F}{\partial P_1^2})^{-1} = \frac{1}{2\{a_1^* + 6a_{11}^* P_1^2 + a_{12}^* P_2^2 + a_{13}^* P_3^2 + 15\alpha_{111} P_1^4 + \alpha_{112}[6P_1^2(P_2^2 + P_3^2) + (P_2^4 + P_3^4)] + \alpha_{123} P_2^2 P_3^2\}},$$

$$\varepsilon_{22} = (\frac{\partial^2 F}{\partial P_2^2})^{-1} = \frac{1}{2\{a_2^* + 6a_{11}^* P_2^2 + a_{12}^* P_1^2 + a_{13}^* P_3^2 + 15\alpha_{111} P_2^4 + \alpha_{112}[(P_1^4 + P_3^4) + 6P_2^2(P_1^2 + P_3^2)] + \alpha_{123} P_1^2 P_3^2\}},$$

$$\varepsilon_{33} = (\frac{\partial^2 F}{\partial P_3^2})^{-1} = \frac{1}{2\{a_3^* + 6a_{33}^* P_3^2 + a_{13}^* (P_1^2 + P_2^2) + 15\alpha_{111} P_3^4 + \alpha_{112}[(P_1^4 + P_2^4) + 6P_3^2(P_1^2 + P_2^2)] + \alpha_{123} P_1^2 P_2^2\}}.$$

The pyroelectric response $p$ along the z-axis direction is given as the summation of the variation of the spontaneous polarization and the dielectric permittivity $\varepsilon$ along z-axis with the temperature

$$p = \left(\frac{\partial D}{\partial T}\right)_E = \frac{\partial P_s}{\partial T} + E\frac{\partial \varepsilon_{33}}{\partial T}, \qquad (12)$$

where $D$ is the dielectric displacement, $P_s$ is the out-of-plane polarization in the absence of the applied field.

### III. Results and Discussions

In this section, we consider the effects of anisotropic in-plane strains on the thermodynamic properties of epitaxial BST50/50 thin films at room temperature $T = 25°C$.

To determine equilibrium thermodynamic states of the films, we calculate all of the minima of $F$ with respect to the components of the polarization and then select the phase that corresponds to free energy minimum. The misfit strain-misfit strain phase diagram of single-domain BST50/50 thin films at room temperature (RT) is shown in Fig. 1. The $\gamma$ phase does not exist in the entire region of misfit strains ranging from –0.006 to 0.006. The other six phase, i.e., three orthorhombic phases ($a_1a_2, a_1c, a_2c$) and three tetragonal phases ($c, a_1, a_2$) are shown in different regions of misfit strains. The out-of-plane polarization $P_3 = 0$ above the dashed line, while $P_3 \neq 0$ under the dashed line. Compared to BT films with anisotropic in-plane strains [13], there exists a new paraelectric phase in the phase diagram of BST50/50 films. As BT is a typical ferroelectric material and its Curie temperature is $120°C$ far away from RT, it stays in ferroelectric state in the whole range. While SrTiO$_3$ is known as an intrinsic quantum paraelectric [21,22]. With doped Sr$^{2+}$, Curie temperature of BST50/50 is $-23°C$ just below RT. It is the doped Sr$^{2+}$ that results in the appearance of the paraelectric phase in the BST phase diagram. With the new phase, it has been chosen as the material for use in many electric devices. $a_1c$ and $a_2c$ orthorhombic phases in BST film appear only in the regions where the anisotropic misfit strains are tensile in one direction and compressive in another direction. Another orthorhombic $a_1a_2$ phase only exists when the misfit strains are tensile in both directions.

We plot the spontaneous polarization as a function of the misfit strain $e_2$ for BST50/50 thin films in the absence of an electric field at RT for fixed value of $e_1 = -0.001$. Fig. 2 shows that with the increase of the misfit strain $e_2$, the $P_3$ component decreases to zero and $P_1$ increases from zero to nonzero values continuously. $P_2$ keeps zero in the whole region from $e_2 = -0.004$ to $e_2 = 0.005$ for $e_1 = -0.001$. This is due to the fact that the large compressive stresses on both directions in-plane shorten the in-plane lattice cell and elongate the z-axis, leading to an increase of out-of-plane polarization. When y-axis direction is the large tensile stress, polarization in this direction increases, and the total polarization is from out-of-plane to in-plane. It can be seen from Fig. 1 that the BST system experiences the $c$ phase, paraelectric phase and $a_2$ phase when misfit strain $e_2$ varies from negative to positive value respectively. We can conclude that the peak value of $\varepsilon_{33}$ occurs where $e_2$ is compressive strain and z-axis is elongate, the peak value $\varepsilon_{22}$ occurs where $e_2$ is tensile strain and y-axis is elongate. Comparing Fig. 2(b) with Fig. 1, the divergence of dielectric constants appears at $e_2 = -0.001$ corresponds to the phase boundary between $c$ phase and paraelectric phase, and at $e_2 = 0.001$ corresponds to the phase boundary between paraelectric phase and $a_2$ phase. $\varepsilon_{11}$ is always flat because no other phase transitions occur at $e_2 = 0.001$.

Fig. 3 shows the effect of misfit strains on the pyroelectric coefficient of BST50/50 thin films in the absence of an electric field at RT. The effect of the applied field on the pyroelectric coefficient of BST60/40 has been investigated in ref. [23]: it shows that the field has little effect on the position of pyroelectric peak. So we consider the case without applied field here. It can be seen that the misfit strains have remarkable effects on the pyroelectric coefficient of BST ferroelectric thin films. We only consider the pyroelectric coefficient along z-axis, which can be seen from eq. (12), thus there's only one pyroelectric peak which corresponding to the phase transition at dashed line in Fig. 1. With the increase of strain $e_1$, the peak of pyroelectric coefficient shift from tensile stress to compressive stress. A maximum pyroelectric coefficient is predicted at the critical misfit strain $e_2$ for fixed value of $e_1 = -0.005, -0.002, 0.002$, and they are corresponding to the transition between $a_2c$ phase and $a_2$ phase, $c$ phase and paraelectric phase, $a_1c$ phase and $a_1$ phase respectively.

In summary, we apply LD phenomenological theory to investigate the anisotropic in-plane strains effects on epitaxial BST50/50 thin films. The anisotropic in-plane strains which are caused by anisotropy lattice constant in x-y plane of BST film or its substrate have great effects on phase transition, dielectric response, and pyroelectric response. Comparing with BT thin films, a new paraelectric phase appears in the diagram of BST50/50 thin film.

**Acknowledgements**

This work was supported by the Open Project of Jiangsu Laboratory of Advanced Functional Materials under the Grant No. 05KFJJ008 and No. 06KFJJ012.

Table 1. Parameters for the calculation of renormalized coefficients for BST50/50 films.

| Parameters | BST50/50 |
|---|---|
| Curie temperature $T_C$ (℃) | -23 |
| Curie constant $C(10^5$ ℃ ) | 1.15 |
| $a_{11}(10^6$ m$^5$/C$^2$F) | 1.87T+740 (T in ) |
| $a_{12}(10^8$ m$^5$/C$^2$F) | 8.75 |
| $S_{11}(10^{-12}$ m$^2$/N) | 4.33 |
| $S_{12}(10^{-12}$ m$^2$/N) | －1.39 |
| $S_{44}(10^{-12}$ m$^2$/N) | 5.01 |
| $Q_{11}$ (m$^4$/C$^2$) | 0.1 |
| $Q_{12}$ (m$^4$/C$^2$) | －0.034 |
| $Q_{44}$ (m$^4$/C$^2$) | 0.029 |

**Figure Captions**

**Fig. 1**: Misfit strain-misfit strain phase diagram of BST50/50 thin films at room temperature.

**Fig. 2**: The spontaneous polarization (a) and relative dielectric constants (b) as a function of the misfit strain $e_2$ for BST50/50 thin films in the absence of an electric field at RT for fixed value of $e_1 = -0.001$.

**Fig. 3:** The pyroelectric coefficient $p$ as a function of misfit strain $e_2$ for BST50/50 thin films in the absence of an electric field at RT.

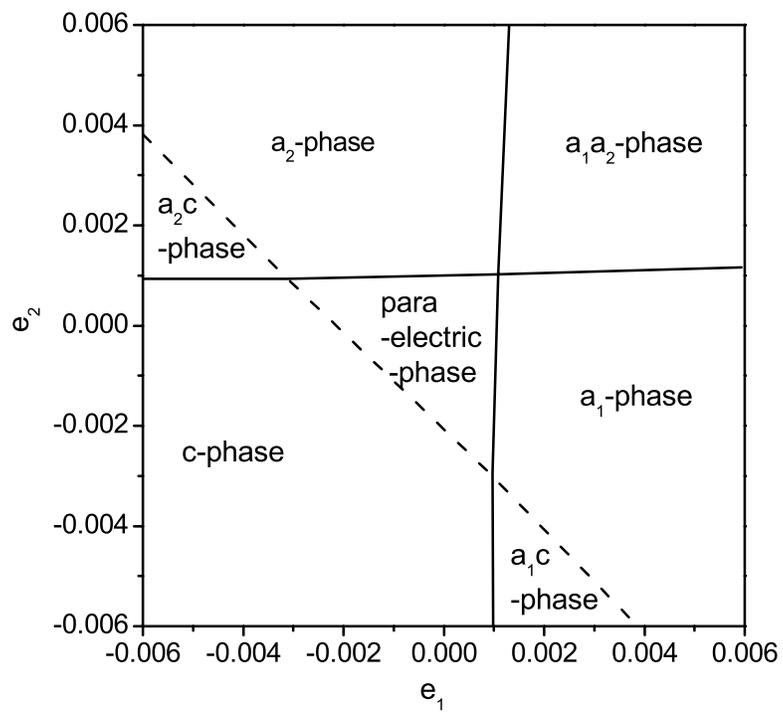

Fig. 1

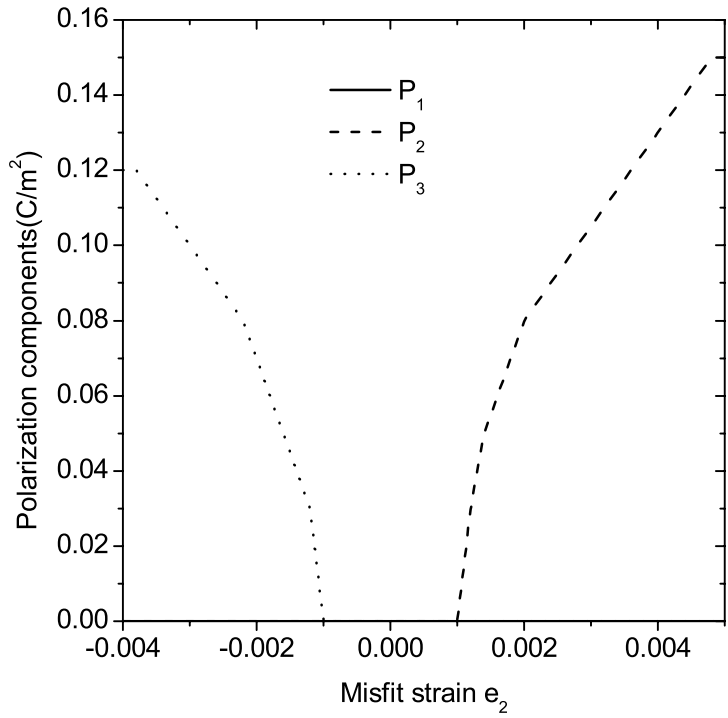

Fig. 2(a)

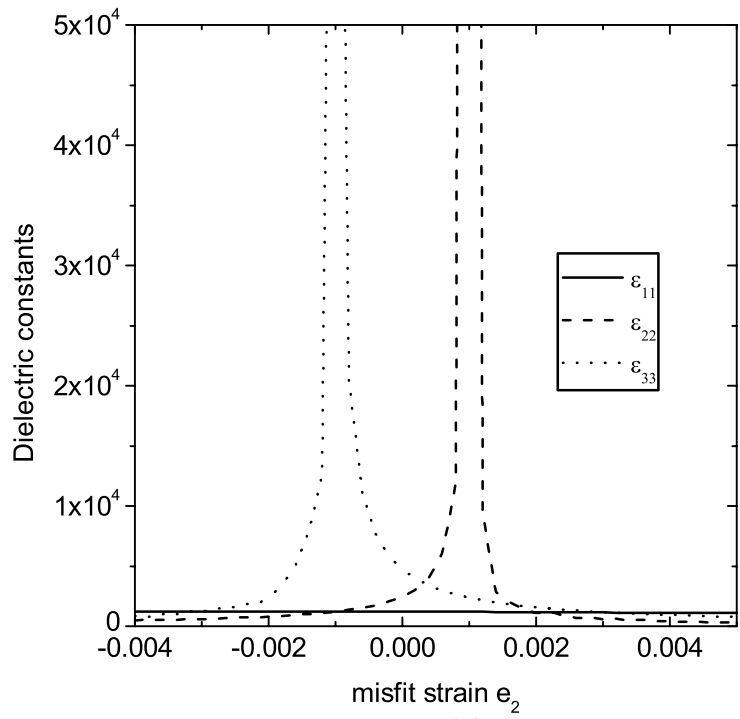

Fig. 2(b)

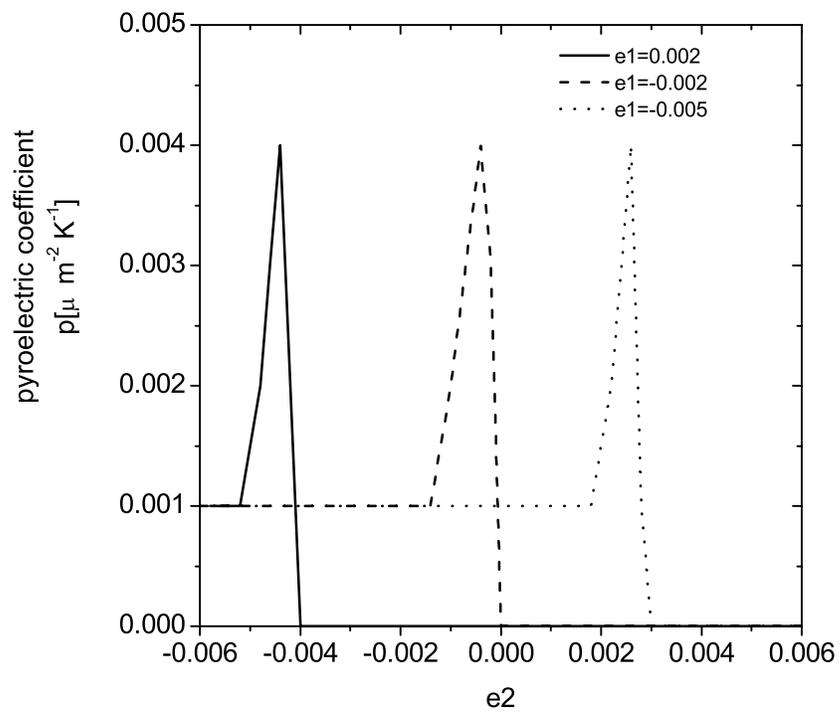

Fig. 3